\documentclass[twocolumn,notitlepage,aps,pra,superscriptaddress,amsmath,amssym,floatfix,longbibliography]{revtex4-1}
\usepackage{subcaption}
\usepackage{graphicx}
\usepackage{bm}
\usepackage{color}
\usepackage{amssymb}
\usepackage{enumerate}
\usepackage{comment}

\def\la{\langle}
\def\ra{\rangle}

\def\be{\begin{equation}}
\def\ee{\end{equation}}

\begin{document}

\newcommand{\bigjprob}{{\mathcal{P}}}
\newcommand{\bigprob}{_{\bm{q}_F}{\mathcal{P}}_{\bm{q}_I}}
\newcommand{\cum}[1]{\llangle #1 \rrangle}       					
\newcommand{\op}[1]{\hat{\bm #1}}                					
\newcommand{\vop}[1]{\vec{\bm #1}}
\newcommand{\opt}[1]{\hat{\tilde{\bm #1}}}
\newcommand{\vopt}[1]{\vec{\tilde{\bm #1}}}
\newcommand{\td}[1]{\tilde{ #1}}
\newcommand{\mean}[1]{\la#1\ra}                  					
\newcommand{\cmean}[2]{ { }_{#1}\mean{#2}}       				
\newcommand{\pssmean}[1]{ { }_{\bm{q}_F}\mean{#1}_{\bm{q}_I}}
\newcommand{\ket}[1]{\vert#1\ra}                 					
\newcommand{\bra}[1]{\la#1\vert}                 					
\newcommand{\ipr}[2]{\left\la#1\mid#2\right\ra}            				
\newcommand{\opr}[2]{\ket{#1}\bra{#2}}           					
\newcommand{\pr}[1]{\opr{#1}{#1}}                					
\newcommand{\Tr}[1]{\text{Tr}(#1)}               					
\newcommand{\Trd}[1]{\text{Tr}_d(#1)}            					
\newcommand{\Trs}[1]{\text{Tr}_s(#1)}            					
\newcommand{\intd}[1]{\int \! \mathrm{d}#1 \,}
\newcommand{\dd}{\mathrm{d}}
\newcommand{\fullint}{\iint \! \mathcal{D}\mathcal{D} \,}
\newcommand{\drv}[1]{\frac{\delta}{\delta #1}}
\newcommand{\partl}[3]{ \frac{\partial^{#3}#1}{ \partial #2^{#3}} }		
\newcommand{\smpartl}[4]{ \left( \frac{\partial^{#3} #1}{ \partial #2^{#3}} \right)_{#4}}
\newcommand{\smpartlmix}[4]{\left( \frac{\partial^2 #1}{\partial #2 \partial #3 } \right)_{#4}}
\newcommand{\limit}[2]{\underset{#1 \rightarrow #2}{\text{lim}} \;}
\newcommand{\funcd}[2]{\frac{\delta #1}{\delta #2}}
\newcommand{\funcdiva}[3]{\frac{\delta #1[#2]}{\delta #2 (#3)}}
\newcommand{\funcdivb}[4]{\frac{\delta #1 (#2(#3))}{\delta #2 (#4)}}
\newcommand{\funcdivc}[3]{\frac{\delta #1}{\delta #2(#3)}}
\definecolor{dgreen}{RGB}{30,130,30}

\title{The Best Radar Ranging Pulse to Resolve Two Reflectors}

\author{Andrew N. Jordan}
\affiliation{The Kennedy Chair in Physics and Institute for Quantum Studies, Chapman University, Orange, CA 92866, USA}
\affiliation{Department of Physics and Astronomy, University of Rochester, Rochester, NY 14627, USA}
\email{jordan@chapman.edu}
\author{John C. Howell}
\affiliation{Institute for Quantum Studies, Chapman University, Orange, CA 92866, USA}
\affiliation{Racah Institute of Physics, The Hebrew University of Jerusalem, Jerusalem, Israel, 91904}
\author{Achim Kempf}
\affiliation{
Department of Applied Mathematics and Department of Physics, University of Waterloo, and
Perimeter Institute for Theoretical Physics, Waterloo, Ontario, Canada}
\author{Shunxing Zhang}
\affiliation{Institute for Quantum Studies, Chapman University, Orange, CA 92866, USA}
\author{Derek White}
\affiliation{Institute for Quantum Studies, Chapman University, Orange, CA 92866, USA}
\date{\today}

\begin{abstract}
Previous work established fundamental bounds on subwavelength resolution for the radar range resolution problem, called superradar [Phys. Rev. Appl. 20, 064046 (2023)].   In this work, we identify the optimal waveforms for distinguishing the range resolution between two reflectors of identical strength.  We discuss both the unnormalized optimal waveform as well as the best square-integrable pulse, and their variants. Using orthogonal function theory, we give an explicit algorithm to optimize the wave pulse in finite time to have the best performance.
We also explore range resolution estimation with unnormalized waveforms with multi-parameter methods to also independently estimate loss and time of arrival. These results are consistent with the earlier single parameter approach of range resolution only and give deeper insight into the ranging estimation problem.  Experimental results are presented using radio pulse reflections inside coaxial cables, showing robust range resolution smaller than a tenth of the inverse bandedge, with uncertainties close to the derived Cram\'er-Rao bound.
\end{abstract}

\maketitle
\section{Introduction} \label{sec:intro}

An outstanding problem in remote sensing is to go beyond established range resolution limits.  The accomplishment of this research task would permit the use of long wavelength waves to resolve features of a target that are many times smaller than the shortest wave component of a pulse.  We have recently demonstrated this possibility in a coaxial cable experiment carrying radio waves \cite{howell2023super}.  In that experiment a bandlimited pulse was used to resolve two depths finer than 10$\times$ the historic radar resolution limits \cite{sheriff1977limitations,skolnik1980introduction,levanon2004radar}.  However, this first experiment did not optimize either the pulse shape, or the data processing of the return signal.  Recent theory calculated classical and quantum Fisher information bounds on the range resolution \cite{jordan2023fundamental}, generalizing the problem to unequal scatterers, as well as multiple scatters for this model system.

The purpose of this article is to find the best radar ranging pulse possible, given the constraint of a fixed band-edge for the Fourier transform.  While we explored some good candidates for different types of waveforms in Ref.~\cite{jordan2023fundamental}, it is of fundamental interest to know the best case result.  We will show this optimal pulse maximizes the Fisher information for the simplest case of the two point reflectors of equal amplitude.  From this point of view, our solution is analogous to the `NOON' states of the best phase estimation states in a quantum optical interferometer \cite{lee2002quantum}, or the GHZ states \cite{greenberger1989going} in quantum metrology theory, used to achieve Heisenberg scaling \cite{giovannetti2011advances}. 

The article is organized as follows:
In Sec.~\ref{sec:range}
we introduce the problem of range resolution in remote sensing.
In Sec.~\ref{sec:field}
we review our previous results in the Fisher information bounds on the range resolution parameter.
In Sec.~\ref{sec:bestwave}, we apply quantum metrology methods to maximize the Fisher information and find the optimal wave.  We continue this analysis in Sec.~\ref{sec:bestpulse} and find methods to produce a finite energy waveform that can  asymptotically reach the Fisher information bound.  In Sec.~
\ref{sec:shift} we show how these solutions can be mapped to arbitrary frequency band intervals.
We reconsider the estimation problem in Sec.~\ref{sec:unnorm} and show how to estimate the time of arrival (total range) as well as the loss and range resolution parameter.  Our results there complement the single parameter results derived previously, and give further insight.  We conclude in Sec.~\ref{sec:conc}.

\section{Range Resolution in Remote Sensing} \label{sec:range}

We consider the simplest case of two equal-amplitude point scatterers of electromagnetic radiation, and our task is to determine ($i$) the minimal discrimination distance of the two targets and ($ii$) the precision on the range between them, focusing on subwavelength range resolution.  It must be stressed that range resolution is a very different task than ranging accuracy - in resolving two or more targets, the reflected waves interfere giving rise to ambiguity in the return signal. We introduce a simple one dimensional model, where the electric field spatial envelope, of the form $f(x)$, is sent out and a return wave of the form 
\be
f_l(x) = \frac{1}{2} (f(x-l/2) + f(x+l/2) ), \label{fl}
\ee
is detected.  For simplicity, we have chosen to set the origin halfway between the two scattering centers.  Here $l$ is the distance between the two scatterers that we wish to estimate.
In a remote sensing context, the amplitude of the returning pulse is typically attenuated from the outgoing intensity by many orders of magnitude, so the absolute amplitude of the pulse is assumed to be unrelated to the target properties, and cannot be used in the range resolution estimation task.  Consequently, we consider the {\it normalized} version of Eq.~(\ref{fl}) and take either the amplitude of the returning field relative to the detector noise, or the number of {\it detected} photons as the metrological resource.  We assume the function $f$ is analytic and normalized, except in Sec.~\ref{sec:unnorm}, where we broaden the estimation problem to also consider time-of-flight and total loss, as well as the range resolution parameter.  We are most interested in the case where the function is {\it band-limited}, having an upper frequency cut-off $f_0$, and when the range resolution parameter, $l < 1/(2\pi f_0)$, is breaking the long-standing trade-off between target resolution and wave carrier frequency \cite{sheriff1977limitations,skolnik1980introduction,levanon2004radar}.  We note that we have set the velocity of the pulse inside the medium equal to unity.  

\section{Fisher information for Field Detection} \label{sec:field}
  The inverse Fisher information bounds the variance of any unbiased estimator $\hat l$ of the parameter $l$, for large data sets
\be
{\rm Var}[{\hat l}] \ge \frac{1}{M I(l)},
\label{crbound}
\ee
where $M$ is the number of repetitions of the measurement, the Cram\'er-Rao bound \cite{cramer1999mathematical}. The Fisher information formalism has been widely applied to bound estimation precision in coherent and incoherent optics, see e.g. \cite{jordan2014technical,motka2016optical,tsang2016quantum}.

We previously found \cite{jordan2023fundamental} that the Fisher information for noisy field detection is given by
\be
I_f \approx \frac{1}{\Sigma^2} \int dx \left(\frac{\partial ({\cal N}_l f_l(x))}{\partial l}\right)^2,  \label{fisher-l}
\ee
where ${\cal N}_l$ is the parameter-dependent normalization, and $\Sigma^2$ is the detector noise power, relative to the signal size.
This result can be further simplified for the range resolution problem as
\be
\Sigma^2 I_f = -(\partial_l \ln {\cal N}_l)^2 + \frac{{\cal N}_l^2}{16} \int dx (f'(x+l/2) - f'(x-l/2))^2.  \label{fieldfi}
\ee
The small $l$ (deep subwavelength) behavior is given by
\be
\Sigma^2 I_f \approx \frac{l^2}{16} \left(  \int dx f''(x)^2 - \left(\int dx f'(x)^2 \right)^2 \right). \label{quad-approx}
\ee
Here we assume well-defined first and second derivatives of the function $f(x)$, as well as square integrability.  The term in parenthesis may be written as
\be
 {\rm Var}\left[ {\hat p}^2 \right]_f =   \int dx f''(x)^2 - \left(\int dx f'(x)^2 \right)^2, \label{varp2}
\ee
where we defined the variance of the operator ${\hat p} = -i \partial_x$ in the state $f$ as ${\rm Var}[...]_f$.
We see immediately that in the position basis, we recover our previous result for the classical Fisher information, as is expected.

The variance of ${\hat p}^2$ can be calculated most easily in momentum space $(k)$ to find
\be
{\rm Var }\left[ {\hat p}^2 \right]_f = \int_{-\infty}^{\infty} dk |{\tilde f}(k)|^2 k^4 -  \left(\int_{-\infty}^{\infty} dk |{\tilde f}(k)|^2 k^2 \right)^2, \label{var-p}
\ee
where ${\tilde f}$ is the Fourier transform of $f$.
Band limited functions have a spectral weight that is exactly zero beyond the band edges $[- k_0, k_0 ]$. 

\section{The best wave} \label{sec:bestwave}
From the Cram\'er-Rao bound, we get the best precision on the range resolution when the Fisher information is maximized.  Focusing on the deep subwavelength resolution case, we seek waveforms that maximize the variance of the square of the momentum, given that we have a constraint that the pulse is band-limited.  That is, the spectral weight is exactly zero for $|k| > k_0$.  We can leverage results in quantum metrology \cite{davidovich2018towards} because we are dealing with normalized waveforms.
It is well-known in the quantum metrology literature that to maximize the variance of an operator ${\hat A}$, the state is prepared such that it is an equal superposition between the maximum and minimum eigenvalue,
\be
|\psi\ra = \frac{|a_{\rm max}\ra + |a_{\rm min}\ra}{\sqrt{2}},
\ee
where we introduced the eigenstates $|a_j\ra$ of the operator ${\hat A}$ such that ${\hat A} |a_j\ra  = a_j |a_j\ra$. The values $a_{\rm max}$ or $a_{\rm min}$ are the maximum or minimum eigenvalue.

Straightforward calculation shows that the variance of $\hat A$ in this state gives
\be
\la \psi | {\hat A}^2 | \psi \ra - 
\la \psi | {\hat A} | \psi \ra^2 =  
\frac{1}{4} (a_{\rm max} - a_{\rm min})^2. \label{var}
\ee
Applied to our case, we are interested in ${\hat A} = {\hat p}^2$. For band-limited functions the maximum eigenvalue is $k_0^2$, which can be realized with momentum eigenstates $|\pm k_0 \ra$.  We take the symmetric combination,
$|\lambda_{\rm max}\ra = (| k_0 \ra + | -k_0 \ra)/\sqrt{2}$.
The minimum eigenvalue is 0, corresponding to the momentum eigenstate $| \lambda_{\rm min}\ra= |0\ra$, corresponding to a dc off-set.

Therefore the optimal state is given by $|\psi\ra =  (1/2) (|k_0\ra + |-k_0\ra) + (1/\sqrt{2}) |0\ra$.  This corresponds to a 3 tooth frequency comb with weights 1/2 at the band edges and weight $1/\sqrt{2}$ at zero frequency.  In real space the wave is given by
\be
\psi(x) = \cos(k_0 x) + \frac{1}{\sqrt{2}}. \label{optwave}
\ee
The overall amplitude is unimportant since this is a non-normalizable wave.  The variance of the momentum squared for the wave is given from Eq.~(\ref{var}) by 
\be
{\rm Var}[{\hat p}^2]_\psi = k_0^4/4. \label{varp2}
\ee
This result sets the upper bound of the variance for band-limited waves.
Unfortunately, the wave exists across all space, and also has a DC off-set, making it useless for radar ranging.  In the next section, we will show we can approach this performance using a finite-energy pulse.

The reason the DC offset is needed is that the height of the two comb-teeth at the band edges changes when the separation $l$ is varied, but the overall signal return is assumed to be unrelated to the scattering problem.  Similar to the method used in our previous article \cite{howell2023super}, the DC offset gives a {\it self-referencing feature} to the pulse - the zero frequency comb tooth is insensitive to the separation $l$, so its height, compared to the band-edge heights, permits the unambiguous estimation of the separation of the reflectors.

\section{The best pulse}
\label{sec:bestpulse}
While the above result is optimized for a pure sinusodial wave, we often require that only normalizable functions are permitted that vanish sufficiently fast for large time (or space), corresponding to a finite energy solution.

We begin by rescaling the momentum scale in units of $k_0$, $(p = k_0 u)$ so that the scaled variance of ${\hat p}^2$ is given by
\be
\frac{4 {\rm Var}({\hat p}^2)}{k_0^4} = 4\left[ \int_{-1}^1 du |{\tilde f}(u)|^2 u^4 - \left( \int_{-1}^1 du |{\tilde f}(u)|^2 u^2\right)^2\right]. 
\ee
According to the result (\ref{varp2}) of the previous section, this quantity, defined as ${\cal R} = 4 {\rm Var}({\hat p}^2)/k_0^4$, must be between [0, 1].

\begin{figure}
    \centering
\includegraphics[width=\columnwidth]{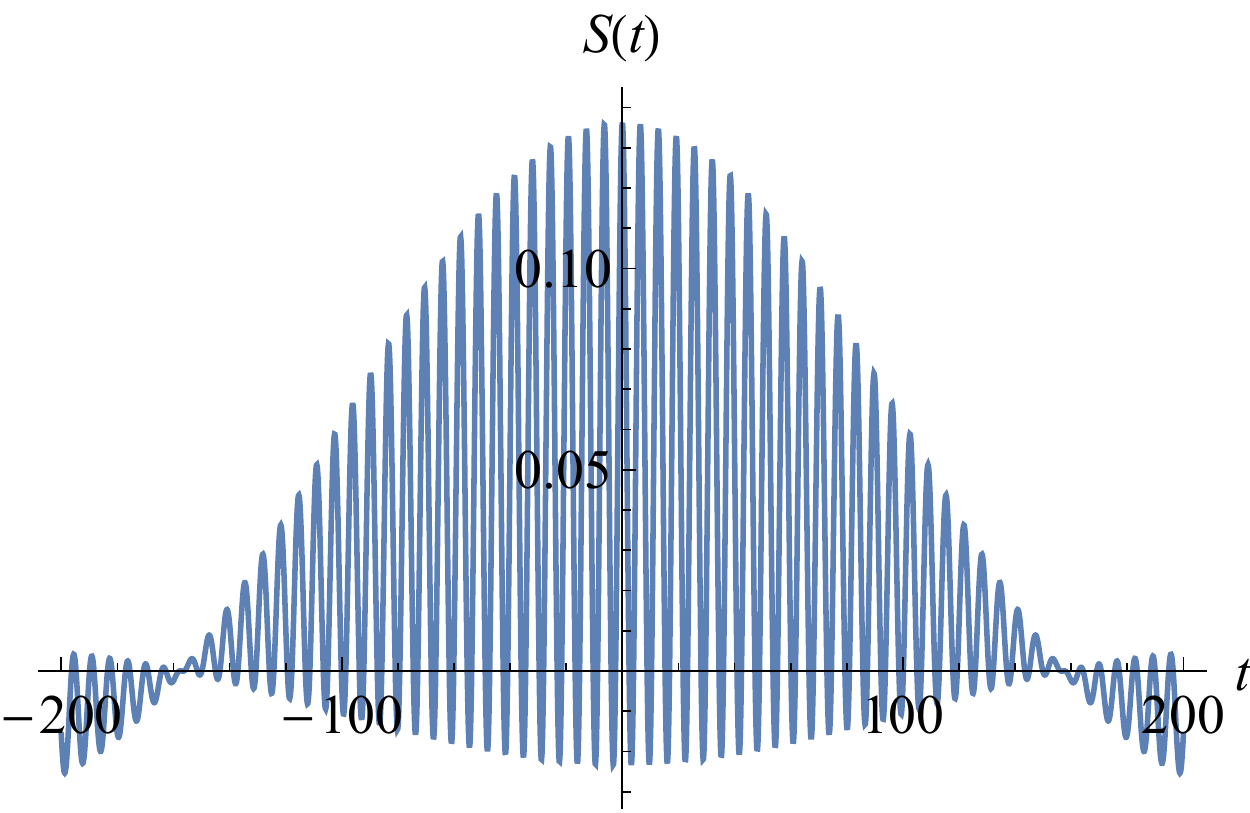}
    \caption{Plot of $S(t)$ versus $t$ (in units of $1/k_0$) for the value $d=50$, corresponding to ${\cal R} \approx 0.92$.}
    \label{fig:howellsinc}
\end{figure}

\begin{figure}
    \centering
\includegraphics[width=0.8\columnwidth]{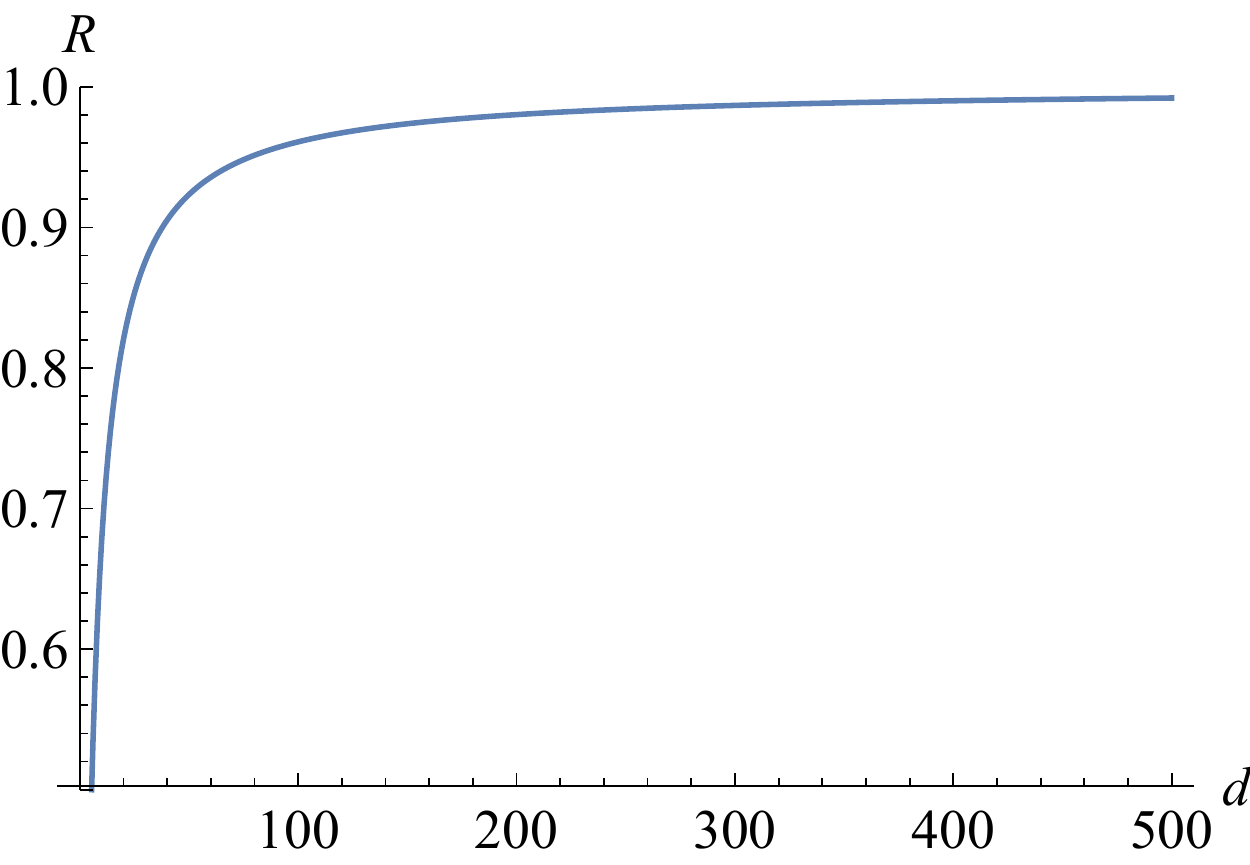}
    \caption{Plot of $\cal R$ versus $d$.}
    \label{fig:ratiosinc}
\end{figure}

One simple way to create a bandlimited waveform is to take the optimal wave (\ref{optwave}) and multiply it by a sinc function.  This will shift the band edge, but we can readjust the wave frequency so the band edge is kept the same:
\be
S(x) = \sqrt{\frac{k_0}{\pi}} {\rm sinc}\left(\frac{k_0 x}{d}\right) \left( \cos \left( k_0 x \left(1-\frac{1}{d}\right)\right) + \frac{1}{\sqrt{2}}\right).
\label{eq-s}
\ee
Here the dimensionless parameter $d$ controls how long the sinc function takes to decay in time, giving a finite-energy wave we have chosen to normalize. An example is plotted in Fig.~\ref{fig:howellsinc} for the value $d=50$.  The sinc function has the effect of turning the delta-spikes in the Fourier domain into rectangles.
As the parameter $d$ increases, the resolving ability of the pulses increases toward ${\cal R} =1$, but at the price a progressively longer pulse, see Fig.~\ref{fig:ratiosinc}.

Another approach to finding these special waveforms that can approach the upper limit on the variance of the square-momentum is by using the formalism recently introduced in Ref.~\cite{karmakar2023beyond}, based on orthogonal function theory.
We write the waveform in the Fourier space as a superposition of Legendre polynomials, $P_n(u)$.  
\be
{\tilde f}(u) = \sum_{n=0}^\infty c_n \sqrt{\frac{2n+1}{2}} P_n(u).
\ee
Here $c_n$ are an arbitrary set of complex coefficients. The polynomials are defined on [-1, 1] and define an orthonormal function basis of band-limited functions,
\be
\int_{-1}^1 du \sqrt{\frac{2n+1}{2}} P_n(u) \sqrt{\frac{2m+1}{2}} P_m(u) = \delta_{n,m}.
\ee
We impose the normalization of the pulse,
\be
\int_{-1}^1 du |{\tilde f}(u)|^2 = \sum_{n=0}^\infty |c_n|^2 = 1,
\ee
where we used the orthonormality of the scaled Legendre polynomials.
In the real space, the Fourier transform of the Legendre polynomials are spherical Bessel functions $j_n(t)$.
\be
f(t) = \sqrt{\frac{2}{\pi}} \sum_{n=0}^\infty c_n \sqrt{\frac{2n+1}{2}} i^n j_n(t).
\label{eq-f}
\ee

\begin{figure}
    \centering
\includegraphics[width=\columnwidth]{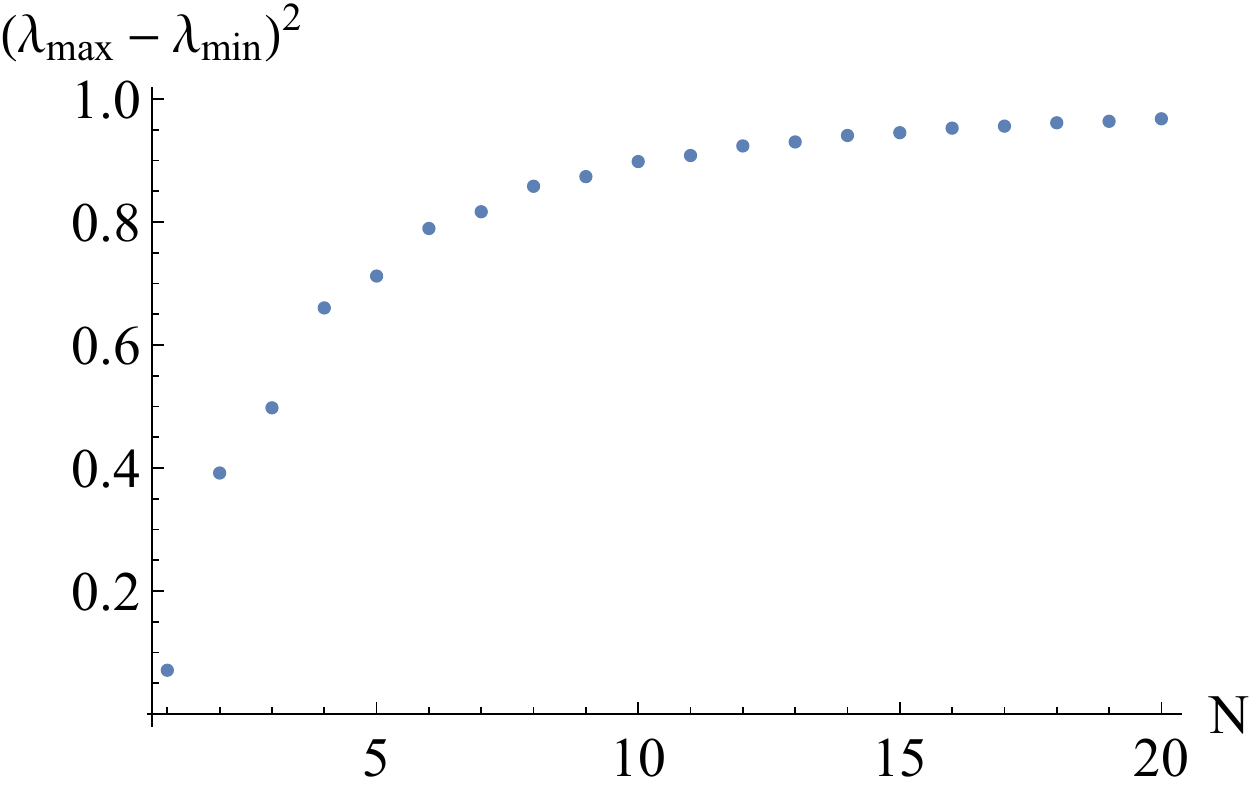}
    \caption{The square different of maximum and minimum eigenvalues of matrix $\bf u^2$ is plotted versus matrix dimension $N$.  It rapidly approaches its upper bound of 1.}
    \label{fig:evaldiff}
\end{figure}

\begin{figure}
    \centering
\includegraphics[width=\columnwidth]{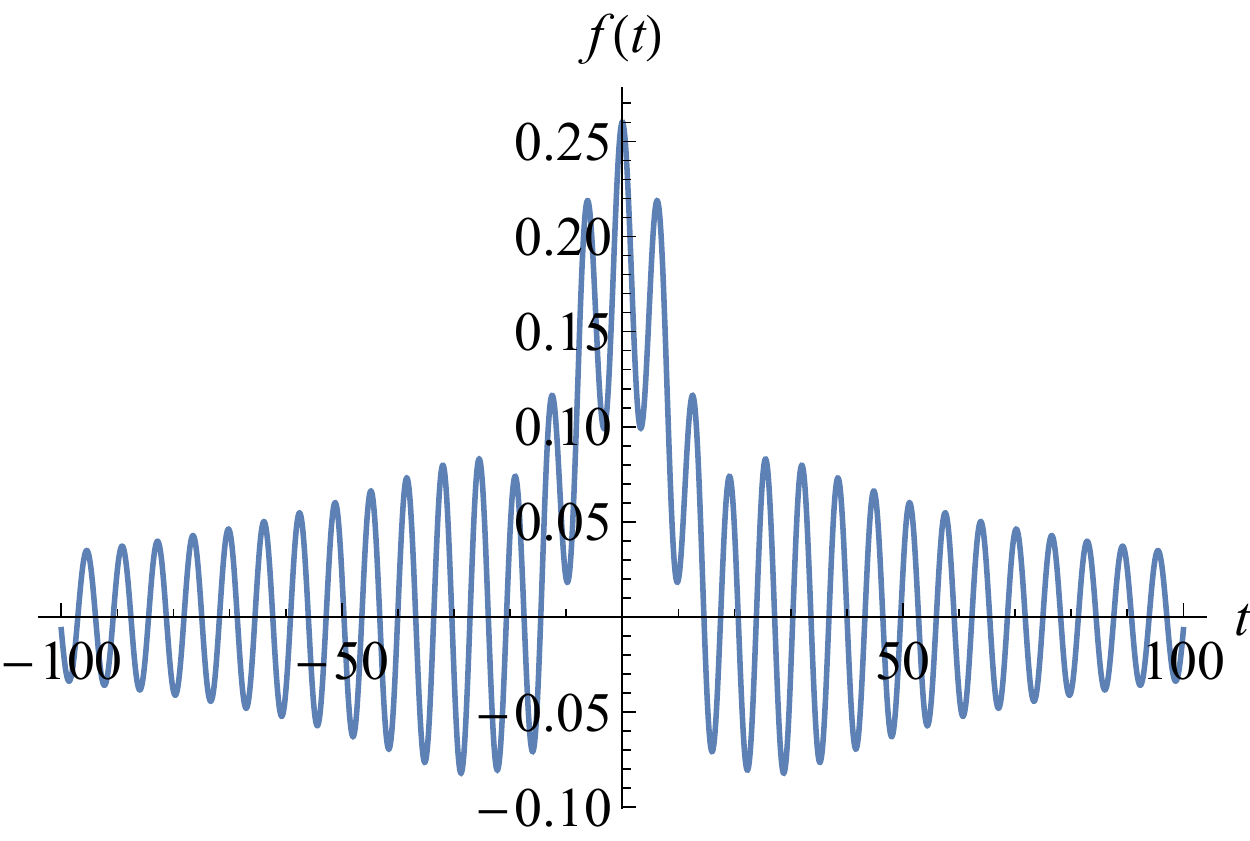}
    \caption{The optimal pulse for the dimension $N=12$ is plotted versus time $t$ (or position), in units of the inverse bandedge.  This function corresponds to ${\cal R} \approx 0.92$.}
    \label{fig:optpulse}
\end{figure}

The second moment of $u$ can be found by
\be
\la u^2 \ra_f = \int_{-1}^1 du |{\tilde f}(u)|^2 u^2 = \sum_{n,m=0}^\infty c_n c^\ast_m (u^2)_{n,m}.
\ee
We have defined the matrix elements
\be
(u^2)_{n,m} = \sqrt{\frac{2n+1}{2}} \sqrt{\frac{2m+1}{2}} \int_{-1}^1 du P_n(u) P_m(u) u^2.
\ee
Therefore the second moment may be written as a quadratic form using a vector of coefficients ${\vec c}$ and a matrix ${\bf u}^2$,
\be
\la u^2 \ra_f  = {\vec c}\,^\dagger \cdot {\bf u^2} \cdot {\vec c}.
\ee
Furthermore, we can diagonalize the matrix ${\bf u^2}$, by introducing eigenvalues $\lambda_n$ and eigenvectors ${\vec v}_n$,
\be
{\bf u^2} \cdot {\vec v}_n = \lambda_n {\vec v}_n.
\ee
We choose the optimized set of coefficients as discussed above in order to maximize the variance of ${\bf u}^2$,
\be
{\vec c}_{\rm opt} = \frac{1}{\sqrt{2}} \left({\vec v}_{\rm max} + {\vec v}_{\rm min}\right),
\ee
in order to maximize the variance of $(\hat p)^2$.  Here the max and min refer to the eigenvectors associated with the smallest and largest eigenvalues.
We then have
\be
{\rm Var}[u^2]_f = \frac{1}{4} (\lambda_{\rm max} - \lambda_{\rm min})^2.
\ee
In practice, we truncate the matrix ${\bf u}^2$ at a finite dimension $N$.  We can then see how the variance of the second moment of the momentum increases as $N$ is increased.  The square difference of the maximum and minimum eigenvalue is plotted versus matrix dimension $N$ in Fig.~\ref{fig:evaldiff}.
For the case where $N=12$, the optimal waveform is plotted in Fig.~\ref{fig:optpulse}.  In both this case, and the example of Fig.~\ref{fig:howellsinc}, the parameter ${\cal R} \approx 0.92$, showing this construction is more compact in time (or space). 

\section{Shifted bandwidth} \label{sec:shift}
Often for technical reasons, we are restricted to work between a finite band $[k_1, k_2]$, that is asymmetric around 0 frequency (we take $k_1, k_2>0$ for simplicity, otherwise we keep the dc offset).  The preceding derivation can be easily adapted to this case.  For the best wave, we simply have a superposition of the upper and lower band edges, 
\be
|\psi \ra = \frac{1}{\sqrt{2} } (|k_1\ra + |k_2\ra),
\ee
so the maximum variance is given by
\be
{\rm Var}[{\hat p}^2]_\psi 
= \Delta k^2\, {\bar k}^2,
\ee
where $\Delta k = k_2 - k_1$ is the bandwidth, and ${\bar k} = (k_1+k_2)/2$ is the central frequency.

To adapt the best pulse results to an asymmetric bandwidth, we quote our earlier results from the general theory of bandlimited functions \cite{karmakar2023beyond} which states that any bandlimited function can always be mapped to the frequency interval $[-1,1]$ by a change of variable.  To map a function $f(\tau)$ of dimensionless time $\tau$ back to a function $g(t)$ of real time and an arbitrary bandwidth $[k_1, k_2]$, we have the correspondence,
\be
g(t) =f\left( \frac{t \Delta k}{2}\right) \exp(i {\bar k} t).
\ee

\begin{figure}
    \centering
    \includegraphics[width=\columnwidth]{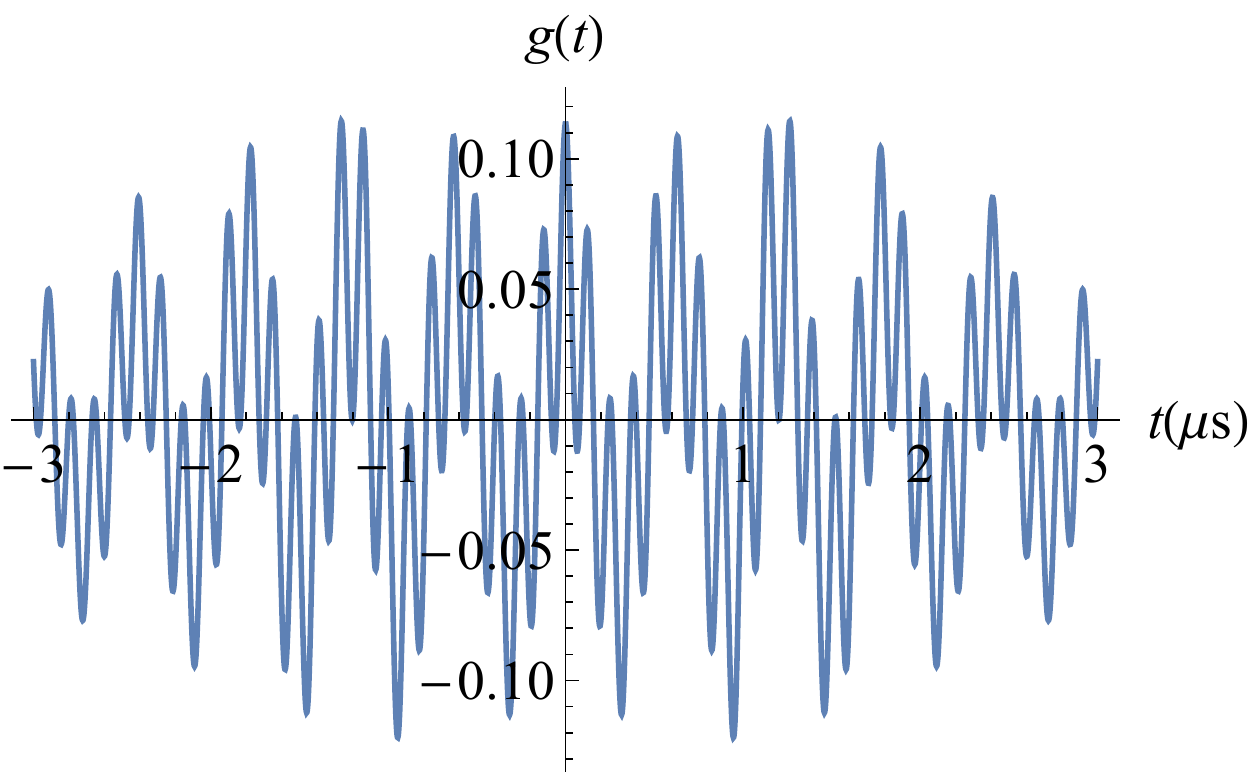}
    \caption{Example of an optimal pulse plotted in time (in $\mu s$) for dimension $N=12$, and a shifted band-limit of $k_1 = (10/2\pi)$ MHz and $k_2= (50/2\pi)$ MHz.}
    \label{fig:shiftband}
\end{figure}

However, in order to find a suitable pulse approximation to this optimal wave, we can apply the same method as before, and instead optimize the variance of ${\hat p}$, which will produce an equally weighted state at the band edges before it is shifted as shown above.  An example of a shifted solution is shown in Fig.~\ref{fig:shiftband}, for a physical band limit of $([10-50]/2\pi)$ MHz.  Before shifting it, the state corresponds to a variance Var$[{\hat p}]_g = \la u^2\ra \approx 0.97$, approaching the upper bound of 1.  The resulting waveform gives a normalized, band-limited, pulsed version of the wave $\cos(10 t) + \cos(50 t)$, where $t$ is here measured in $\mu$s.

\section{Unnormalized waveforms}
\label{sec:unnorm}
It is also possible to work with unnormalized waveforms at the price of also estimating the overall amplitude of the returning signal.
We consider the model where a waveform $f(x)$ is sent out and a returning waveform is measured together with the detector noise $\xi(x)$ with noise power $\sigma^2$ as
\be
s(x_k) = A [f(x-x_0+l/2) + f(x-x_0 - l/2)]/2 + \xi(x).
\ee
Here $A$ is the overall reduction of amplitude (typically orders of magnitude lower than the sent amplitude) that we assume is independent of the other parameters.  The range resolution is $l$.  We also include the temporal (converted to spatial) offset of the pulse $x_0$ that we should also estimate.

Following the treatment of Ref.~\cite{jordan2023fundamental}, we can apply multi-parameter estimation to this model, given data $s(x)$.    The Fisher information for a multi-dimensional Gaussian distribution is 
\be
I_{ij} = \frac{1}{\sigma^2} \sum_k \frac{\partial \mu_{\vec \theta}(x_k)}{\partial \theta_i} \frac{\partial \mu_{\vec \theta}(x_k)}{\partial \theta_j} ,
\ee
where $\mu_{\vec \theta}(x) =  A [f(x-x_0+l/2) + f(x-x_0 - l/2)]/2$ is the mean of the distribution at position $x$, where ${\vec \theta} = (A, x_0, l)^T$ is the parameter vector to be estimated.

We are interested in the deep subwavelength case, so we can expand to leading order in $l$,
\be
\mu_{\vec \theta}(x) \approx  A \left[f(x-x_0) + \frac{l^2}{8} f''(x-x_0)\right].
\ee
A simple example illustrates the challenge:  Consider $f(x-x_0) = \sin(k_0(x-x_0))$.   Then we have $\mu \approx A(1 -k_0^2 l^2/8) \sin(k_0(x-x_0))$.  We see the mean has an effective loss factor $A' = A (1-k_0^2 l^2/8)$ that is impossible to distinguish from the range parameter $l$ if both are fixed for this simplest wave.  We notice that if the wavenumber $k_0$ is changed, that the relative amplitude will change such that $l$ can be estimated - this strategy can be exploited with a frequency comb technique, or even the optimal 3-tooth comb discussed in the previous section.

We write the sum over space in the Fisher information as an integral, $\Delta x \sum_k = \int dx $, where $\Delta x$ is the discritization of space (converted from time), and define $\Sigma'^2 = \Delta x \sigma^2$ to find the Fisher information (to leading order) to be
\begin{widetext}
\be
\Sigma'^2 {\bf I} = \begin{pmatrix} \int dx f(x-x_0)^2 & 0 & \frac{A l}{4} \int dx f(x-x_0) f''(x-x_0) \\
0 & A^2 \int dx f'(x-x_0)^2 & 0 \\
\frac{A l}{4} \int dx f(x-x_0) f''(x-x_0) & 0 & \frac{A^2 l^2}{16} \int dx f''(x-x_0)^2 \end{pmatrix}.  \label{multi-fi}
\ee
\end{widetext}
From here on we set the normalization of the sent waveform to be 1 for simplicity, $\int dx f^2 = 1$. 
Note that the zero off-diagonal elements involve either $\int dx f(x-x_0) f'(x-x_0)$ or $\int dx f'(x-x_0) f''(x-x_0)$, both of which are integrals of total differentials, which then vanish for finite time pulses.

The arrival time $x_0$ is asymptotically uncorrelated with the amplitude $A$ or range resolution $l$ because of the block form of the Fisher information matrix, so it can be independently estimated.  However, as noted above, the estimation of $l$ independently from $A$ is challenging because there are off-diagonal elements.  Focusing on just the estimation of those two parameters reduces the Fisher information to a 2 by 2 matrix.  Inverting this matrix then gives the matrix form of the Cram\'er-Rao bound,
${\rm Var}[\theta_i \theta_j] \ge ({\bf I}^{-1})_{ij}$, $i, j = A, l$, where the inverse Fisher information matrix is given by
\be
{\bf I}^{-1} = \frac{\Sigma'^2}{ {\rm Var}\left[ {\hat p}^2 \right]_f } \begin{pmatrix}
 \int dx (f'')^2 & \frac{4}{l A} \int dx (f')^2 \\ \frac{4}{l A} \int dx (f')^2 &
 \left(\frac{4}{l A}\right)^2
\end{pmatrix}.
\ee
Here we used the notation introduced in Eq.~(\ref{varp2}).  If we are only interested in the range resolution parameter $l$, it is interesting to see the $({\bf I}^{-1})_{ll}$ element of the inverse Fisher information matrix returns the same result we derived in the deep-subwavelength limit, Eq.~(\ref{quad-approx}).  The noise power $\Sigma^2$ is effectively scaled by $A^2$, so $\Sigma' = A \Sigma$.

We can now apply Maximum Likelihood estimation to this multi-parameter estimation problem.  We note that unlike the treatment in Ref.~\cite{jordan2023fundamental}, where the estimation of the range resolution is done discarding the total power in the pulse, here we estimate the three parameters ${\vec \theta} = (A, x_0, l)^T$ separately.  The estimators are found by considering the maximum likelihood $\partial {\cal L}/\partial {\theta_j}=0$, where the likelihood is given by
\be
{\cal L} = - \frac{1}{2\Sigma'^2}
 \int dx \left(s(x) - \mu_{\vec \theta}(x)\right)^2.
 \ee
The maximization results in the equations $\int dx (s-\mu_{\vec \theta}(x))f^{(n)}(x)=0$ for derivatives $n=0,1,2$.
Replacing the resulting variables with their estimators ${\hat \theta_j}$, gives the optimal estimators.  Solving for the squared range resolution, we find
\be
\frac{{\hat l^2}}{8} = \frac{1}{{\rm Var}[{\hat p}^2]_f} \left( \frac{
\int dx s(x) f''(x)}{\int dx s(x) f(x)} +\int dx (f'(x))^2 \right).
\ee

 For multiple repetitions, we replace the data $s(x)$ by its statistical average at each position $x$. Clearly if the variance of ${\hat p}^2$ vanishes, this solution does not exist, as we anticipated earlier in the section.

\section{Experimental Implementation}

\begin{figure}
    \centering
    \includegraphics[width=\columnwidth]{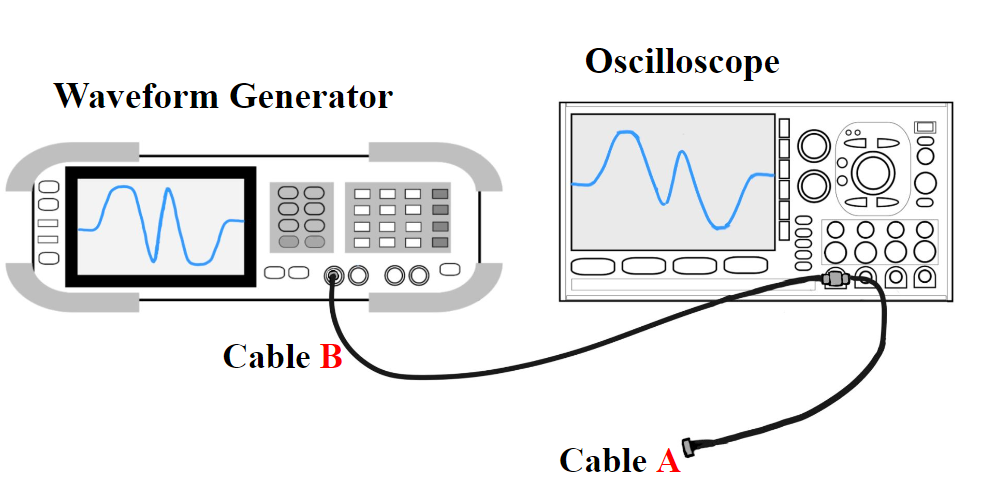}
    \caption{Proof-of-concept experimental setup of the optimal pulse for two scatterers. A waveform generator sends a constructed pulse into a double cable setup with the length of cable A being $l/2$.}
    \label{fig:exp_setup}    
\end{figure}

\begin{figure}
    \subfloat[Plot of the mean of $\hat{l}/V_{\tau}$ vs. $l/V_{\tau}$.  The shaded band indicates the 1 sigma statistical uncertainty in the mean.]{%
    \includegraphics[clip,width=0.9\columnwidth]{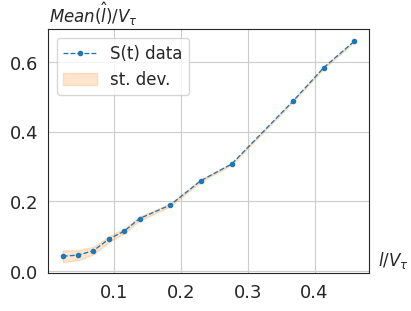}%
    }
    
    \subfloat[Plot of the variance of $\hat{l}/V_{\tau}$ vs. $l/V_{\tau}$]{%
    \includegraphics[clip,width=0.9\columnwidth]{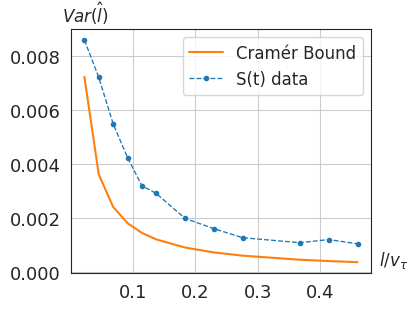}%
    }
    
    \caption{$V_{\tau}$ is the inverse bandedge of the sinc-cosine function based $S(t)$ pulse with $d=50$. The true separation distance $l$ is kept constant at 0.61m, and $V_{\tau}$ is varied. The orange line in plot b) shows the theoretical bound derived from the Cram\'er-Rao lower bound (\ref{crbound}).}
    \label{fig:S(t)_results}
\end{figure}

We put in the specially-designed pulses $S(t)$, Eq.~(\ref{eq-s}) and $f(t)$, Eq.~(\ref{eq-f}) into an arbitrary waveform generator, which then outputs those waveforms into a 50$\Omega$ BNC cable network and is measured by an oscilloscope.  Fig.~\ref{fig:exp_setup} illustrates the schematic representation of the experimental apparatus. Upon exiting the waveform generator, the pulse traveled through Cable B and bifurcated at the T-junction, with a portion propagating through Cable A. Due to impedance mismatches, the partial pulse reflected at the open end of Cable A, traveling a round-trip length of $l$, where $l/2$ is the length of Cable A. In the particular BNC setup shown in Fig. \ref{fig:exp_setup}, two pulses are generated.  The objective is to investigate the resolvability in time (or space) by adjusting $V_{\tau}$ (the inverse bandedge) of the pulse while maintaining a constant length of $l$. 

Multiple pulses were collected from the oscilloscope across various pulse lengths for a single Cable A length of 0.305 m, so the round trip length $l$ is 0.61 m. The composite return pulse waveforms are of the type $c_0 S(t) + c_1 S(t+L/v)$, where $c_0$ and $c_1$ denote the respective amplitudes of the primary pulse and its reflection, and $v$ the speed of the radio wave in the cable. We employed an root mean square error (RMSE) grid search method to determine the optimal parameters for Cable A length ($l$) and pulse amplitudes ($c_0$ and $c_1$). This iterative optimization technique utilized the collected pulse data alongside the base pulse waveform to search through the parameter space until the best fitting parameters are found, thereby achieving the best fit for the observed pulses.

\begin{figure}
    \subfloat[Plot of the mean of $\hat{l}/V_{\tau}$ vs. $l/V_{\tau}$. The shaded band indicates the 1 sigma statistical uncertainty in the mean.]{%
    \includegraphics[clip,width=0.9\columnwidth]{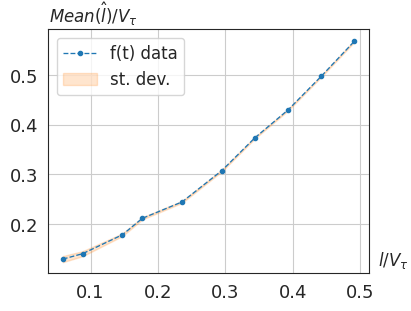}%
    }
    
    \subfloat[Plot of the variance of $\hat{l}/V_{\tau}$ vs. $l/V_{\tau}$]{%
    \includegraphics[clip,width=0.9\columnwidth]{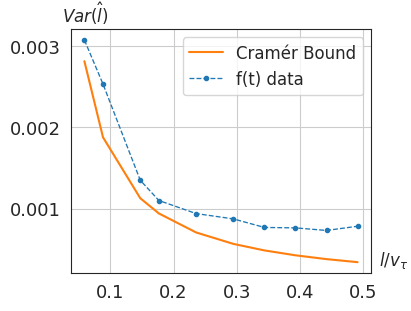}%
    }
    
    \caption{$V_{\tau}$ is the inverse bandedge of the spherical-Bessel function based $f(t)$ pulse with mode dimension $N=12$. The true separation distance $l$ is kept constant at 0.61m, and $V_{\tau}$ is varied. The orange line in plot b) shows the theoretical bound derived from the Cram\'er-Rao lower bound (\ref{crbound}).}
    \label{fig:f(t)_results}
\end{figure}

Consider the time-dependent pulses shown in Figs.~\ref{fig:howellsinc} and \ref{fig:optpulse} and results shown of Figs.~\ref{fig:S(t)_results} and \ref{fig:f(t)_results}. The time-dependent functions $S(t)$ and $f(t)$ are shown in the first two figures, while the variance and mean of $\hat{l}$ are shown in the last two. Specifically, Part (b) of these figures illustrates the variance of the measured $\hat{l}$ against the true separation $l/V_{\tau}$, compared to the Cram\'er-Rao bound. This analysis facilitated the examination of the relationship between the variance in Cable A length ($l/2$) and the normalized length-to-bandedge ratio ($l/V_{\tau}$). The lower the variance of the measured $\hat{l}$ is, the better the resolvability of the relative range. The pulse $f(t)$ was closer to the Fisher information-derived bound than the $S(t)$ pulse. We believe this comes from the more compact pulse arising from the spherical Bessel function method, which is not subject to long-time correlated noise.

Additionally, Part (a) of Figs.~\ref{fig:S(t)_results} and \ref{fig:f(t)_results} shows the mean of the measured $\hat{l}/V_{\tau}$ versus $l/V_{\tau}$. We anticipate a direct 1:1 correlation between the means of the measured $\hat{l}/V_{\tau}$ and $l/V_{\tau}$. However, the plot comparing $\hat{l}/V_{\tau}$ against $l/V_{\tau}$ reveals that the measured $\hat{l}/V_{\tau}$ is marginally greater than $l/V_{\tau}$, deviating larger than the expected statistical uncertainty. This discrepancy, revealing systematic error, is attributed to the experimental setup of the BNC cables (shown in Fig.~\ref{fig:exp_setup}), which does not conform perfectly to a two-scatterer model. Instead, there is a series of small reflections from the T-junction to the cable's dead end, producing an infinite series of exponentially diminishing amplitude reflections. This makes our model give a systematically larger $\hat{l}$ value than the experimental reality, accounting for the mismatch between the mean and true value of the parameter.

\section{Conclusions} \label{sec:conc}

We have considered the problem of the optimal wave and waveform for the range resolution estimation problem in the simplest case of two scatters in a one dimensional geometry.  By optimizing the Fisher information of the range resolution parameter, we found explicit constructions for the optimal solution, given the space of bandlimited functions with a fixed bandedge.  In practice this bandedge is set by the constraints of the environment one is working in, such as the absorption behavior of water versus frequency.  We also reexamined the range resolution problem from the point of view of multi-parameter estimation, where the total loss and timing information of the return pulse is also considered.  Our results were consistent with the simplest single parameter results derived previously.  We have focused here on equal strength parameters, but this methodology can be adapted to unequal strength reflectors as well as multiple reflectors, as was considered in \cite{jordan2023fundamental}.

We have experimentally implemented the optimal pulses experimentally, using both the sinc-envelope optimal wave as the spherical Bessel function based method with matched band edges.  Our experimental results show robust superradar range resolution.  Both pulses are comparable in uncertainty to the theoretical limits set by the Cram\'er-Rao bound.  The spherical Bessel function method performs slightly better.


\section{Acknowledgements}
Support from Chapman University and the Bill Hannon Foundation is gratefully acknowledged.  We thank Luis Sanchez-Soto and  Mankei Tsang
for helpful discussions.  This research was supported in part by grant NSF PHY-1748958 to the Kavli Institute for Theoretical Physics (KITP).

\end{document}